\documentclass[twocolumn]{article}
\usepackage[a4paper, total={180mm, 243mm}]{geometry}
\usepackage[T1]{fontenc}
\usepackage{lipsum}
\usepackage{graphicx}
\usepackage{siunitx}
\usepackage{upgreek}
\usepackage{titling}
\usepackage{circledsteps}
\usepackage{subcaption}
\usepackage[version=4]{mhchem}
\usepackage{amsmath, amssymb}
\usepackage{soul}
\usepackage{empheq}

\definecolor{authorcolor}{cmyk}{0.8,0.6,0,0.3}

\DeclareCaptionLabelSeparator{custom}{\textbar}
\DeclareCaptionFormat{custom}{\textsf{\textbf{#1 #2} \small #3}}
\graphicspath{{./figures}} 

\makeatletter
\newcommand{\showfontsize}{\f@size{} pt}
\makeatother

\usepackage[
    backend=biber,
    style=nature,
    url=false,
    isbn=false,
    date=year,
    maxnames=99
]{biblatex}
\let\cite\supercite

\newcommand{\reffig}[2]{Fig.~\ref{#1}{\textbf{#2}}}

\newcommand\identity{1\kern-0.25em\text{l}}

\title{Asynchronous Terahertz-field sampling with soliton microcombs}

\author{Alexandra Sakharova$^{1\dagger}$,
        Alexander~E.~Ulanov$^{1\dagger}$,
        Konstantin Wenzel$^{2\dagger}$,\\
        Milan Deumer$^{2}$,
        Alexander Dohms$^{2}$,
        Bastian Ruhnke$^{1}$,\\
        Lars Liebermeister$^{2}$,
        Robert B. Kohlhaas$^{2}$,
        Tobias Herr$^{1,3,*}$
}
\date{%
    \small $^1$Deutsches Elektronen-Synchrotron DESY, Notkestr. 85, 22607 Hamburg, Germany \\
    \small $^2$Fraunhofer Institute for Telecommunications, Heinrich Hertz Institute, Einsteinufer 37, 10587 Berlin, Germany\\
    \small $^3$Institute of Experimental Physics, University of Hamburg UHH, Luruper Chaussee 149, 22761 Hamburg, Germany\\
    \small $^{\dagger}$Authors contributed equally
    \small $^{*}$tobias.herr@desy.de
}

\begin{document}
\begin{refsection}

\maketitle

\textbf{
Terahertz time-domain spectroscopy records the electric field of terahertz transients and underpins applications from materials science to non-destructive testing. Asynchronous optical sampling enables systems without moving parts, but its acquisition rate remains fundamentally limited by the repetition-rate of the driving femtosecond lasers. Here, we demonstrate asynchronous terahertz-field sampling with two chip-based soliton microcombs. Their ultrahigh 25~GHz pulse repetition-rate drastically reduces dead-time during acquisition, and stabilization via sideband injection-locking enables coherent averaging without active feedback loops. At 10~kHz acquisition rate, we achieve a dynamic range of 74~dB in a single 100~$\upmu$s scan, and 100~dB after 50~ms of coherent averaging with a 1.6~THz bandwidth. Acquisition rates of up to 100~kHz are demonstrated, limited only by detection electronics. Together with a proof-of-concept silicon wafer thickness measurement, these results establish soliton microcombs as compact sources for rapid terahertz-field sampling.
}

\subsection*{Introduction}
Terahertz time-domain spectroscopy (THz-TDS)~\cite{koch2023TDS} is a key spectroscopic technique in the 0.1--10~THz band. It underpins fundamental research in physics and materials as well as applications~\cite{tonouchi2007CuttingedgeTerahertzTechnology, jepsen2011TerahertzSpectroscopyImaging, naftaly2019IndustrialApplicationsTerahertz, leitenstorfer20232023TerahertzScience}, ranging from non-destructive inspection of coatings~\cite{fitzgerald2005nondestructive, zeitler2007analysis, tao2020non} and security screening~\cite{Federici2005Security, tzydynzhapov2020new, jiang2026detection} to pharmaceuticals~\cite{Zeitler2007Pharma, bawuah2021advances, moradikouchi2023terahertz} and biomedical sensing~\cite{knobloch2002medical, Yu2012Biomedical, cong2023biomedical, yang2016biomedical}. In THz-TDS, a first femtosecond pulse generates a THz transient usually in a photoconductive antenna (PCA) or by optical rectification. A second femtosecond pulse with a variable delay samples the THz field after interaction with the sample, by gating a second PCA or via electro-optic detection. 

\begin{figure*}[!t]
  \centering
  \includegraphics[width=\textwidth]{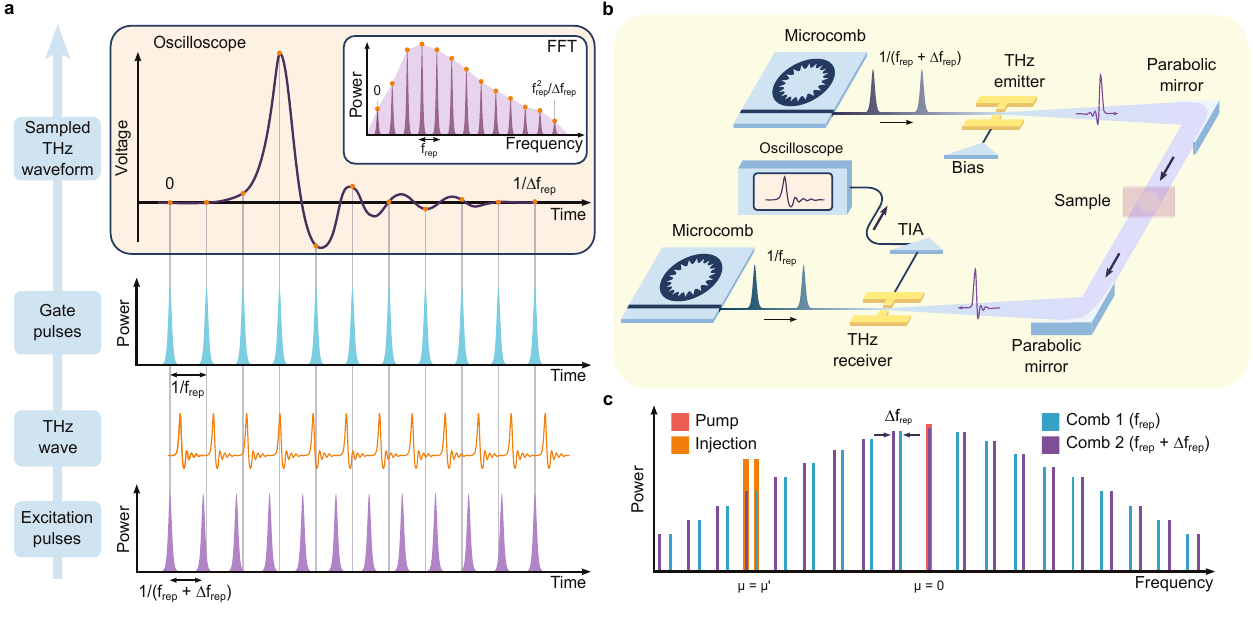}
  \caption{\textbf{Concept of THz-TDS with soliton microcombs.}
    \textbf{a}, Asynchronous optical sampling technique using two optical pulse trains for THz generation (emitter) and THz sampling (receiver) with scan range of $1/\Delta f_\text{rep}$. The sampled THz signal is stretched in time by a factor $f_\text{rep}/\Delta f_\text{rep}$. Inset: Spectrum of the THz waveform obtained via Fast Fourier transform (FFT) of the time domain signal. FFT of a train of sampled THz signals results in a comb-like spectrum.
    \textbf{b}, Conceptual setup for asynchronous THz field sampling with microcombs. Two chip-integrated Si$_3$N$_4$ photonic crystal microresonators (PhCR) serve as ultra-short pulse sources with repetition rate difference $\Delta f_\text{rep}$. Pulses from a first microcomb are sent to an emitter, generating a THz transient which, after collimation, travels through a sample. Pulses from a second microcomb are sent to a photoconductive receiver where they gate the THz transient to generate a momentary field-dependent nA-level current. A transimpedance amplifier (TIA) converts the current to a voltage signal, which is sampled by an oscilloscope. (See \textbf{a}).
    \textbf{c}, Illustration of the microcomb spectra, with a shared continuous-wave pump line at the central mode with relative mode number $\mu=0$. Sideband-injection with two continuous-wave lasers at relative mode $\mu'$ enables all-optical control over their repetition rates $f_\text{rep}$ and $f_\text{rep} + \Delta f_\text{rep}$.
    }
    \label{fig:concept}
\end{figure*}

To accurately sample the THz field, the delay between generating and sampling pulse must be scanned with precisely known timing and low jitter. A single pulse source may be utilized, combined with pulse splitting and a mechanical delay stage. However, mechanical delay stages limit scan rates to the Hz range and introduce vibration and scan nonlinearity. In contrast, asynchronous optical sampling (ASOPS)~\cite{Janke2005ASOPS} uses two femtosecond sources with repetition-rates $f_\mathrm{rep}$ and $f_\mathrm{rep}+\Delta f_\mathrm{rep}$ ($\Delta f_\mathrm{rep} \ll f_\mathrm{rep}$), sweeping the full delay window $1/f_\mathrm{rep}$ at the rate $\Delta f_\mathrm{rep}$ without moving parts (\reffig{fig:concept}{a}). Effectively, ASOPS stretches the THz waveform in time by $f_\mathrm{rep}/\Delta f_\mathrm{rep}$, so that it can be electronically recorded. However, the THz response of condensed-matter systems typically lasts only tens of picoseconds. For well-established sources with $f_\mathrm{rep} \sim 100$ MHz, the resulting 10~ns scan window far exceeds this response duration, so most of each scan consists of \textit{dead-time} carrying no useful signal. Repetition-rate modulation schemes such as OSCAT~\cite{Hochrein2010OSCAT} and ECOPS~\cite{Kim2010ECOPS} confine the sweep to the window of interest and have reached 8~kHz scan rates over 180~ps at the cost of experimental complexity~\cite{Dietz2014ECOPSTDS}.

An elegant way to reduce dead-time is by using higher repetition-rate pulse sources. For condensed-matter samples with a short THz response, $f_\mathrm{rep}$ of tens of GHz is possible without temporal overlap of consecutive responses. Moreover, a higher repetition-rate also permits a faster scan rate ($\Delta f_\mathrm{rep}$), while maintaining the temporal stretching of the THz signal needed for a given electronics bandwidth. Following this route, 1~GHz single-cavity dual-comb lasers have enabled 22~kHz scan rates~\cite{Willenberg2024ThzTDSYb}, and two 10~GHz Ti:sapphire oscillators reached 100~kHz~\cite{Kliebisch2016TDS10GHz100kHz}. Mode-locked lasers at such repetition-rates are, however, demanding to build, operate and stabilize.

Complementing mode-locked lasers, \textit{soliton microcombs} \cite{kippenberg2018DissipativeKerrSolitons,pasquazi2018MicrocombsNovelGeneration, herr2026FrequencyCombsCoherent} readily provide access to femtosecond pulses at high-repetition-rates, spanning from 10 GHz to 1 THz. They are usually generated via the formation of dissipative soliton pulses in continuous-wave driven Kerr-nonlinear microresonators~\cite{leo2010TemporalCavitySolitons, herr2014TemporalSolitonsOptical}. Recent advances have led to electrically-driven chip-scale microcombs with drastically reduced operational complexity~\cite{stern2018BatteryoperatedIntegratedFrequency, shen2020IntegratedTurnkeySoliton, xiang2021LaserSolitonMicrocombs, ulanov2024SyntheticReflectionSelfinjectionlocked, wildi2024PhasestabilisedSelfinjectionlockedMicrocomb}. Microcombs have enabled new applications, including wavelength-multiplexed optical communication \cite{marin-palomo2017MicroresonatorbasedSolitonsMassively}, astronomy \cite{suh2019SearchingExoplanetsUsing, obrzud2019MicrophotonicAstrocomb}, and rapid dual comb spectroscopy~\cite{suh2016MicroresonatorSolitonDualcomb}. Microresonator based systems have also been utilized to generate THz signals~\cite{tetsumoto2021OpticallyReferenced300, Zhang2019peanut, wang2021HighpowerHighcoherenceIntegrated, kuse2022LowPhaseNoise}, including for THz-TDS with a mechanically scanned delay~\cite{peters2026MillimetrewaveCombGenerated}.

Here, we demonstrate asynchronous THz-field sampling with two amplified femtosecond soliton microcombs with 25~GHz repetition-rate (\reffig{fig:concept}{b}), reaching a THz bandwidth of up to 1.6~THz. The 40~ps pulse period exceeds the duration of condensed-matter THz transients and largely eliminates the dead-time of lower-repetition-rate systems. The high pulse repetition-rate also enables a high acquisition rate $\Delta f_\mathrm{rep}$, which we set at 10, 25 and 100~kHz; while the microcombs would support MHz scan rates, the rate here is limited by the sampling electronics. Repetition-rates are stabilized by Kerr-sideband injection locking (Kerr nonlinear synchronization)~\cite{jang2018SynchronizationCoupledOptical, yang2019VernierSpectrometerUsing, moille2023KerrinducedSynchronizationCavity, wildi2023SidebandInjectionLocking, ulanov2025AttosecondlevelSynchronisationChipintegrated}, which provides the ultra-low relative timing jitter required for coherent averaging (\reffig{fig:concept}{c}). At a 10~kHz scan rate we obtain a dynamic range of 74~dB in a single scan and 100~dB after coherent averaging over 50~ms. As an example, we demonstrate phase-resolved transmission spectroscopy of a silicon wafer. These results position soliton microcombs as compact, high-repetition-rate sources for rapid THz-field sampling and spectroscopy.

\begin{figure*}[!t]
  \centering
  \includegraphics[width=\textwidth]{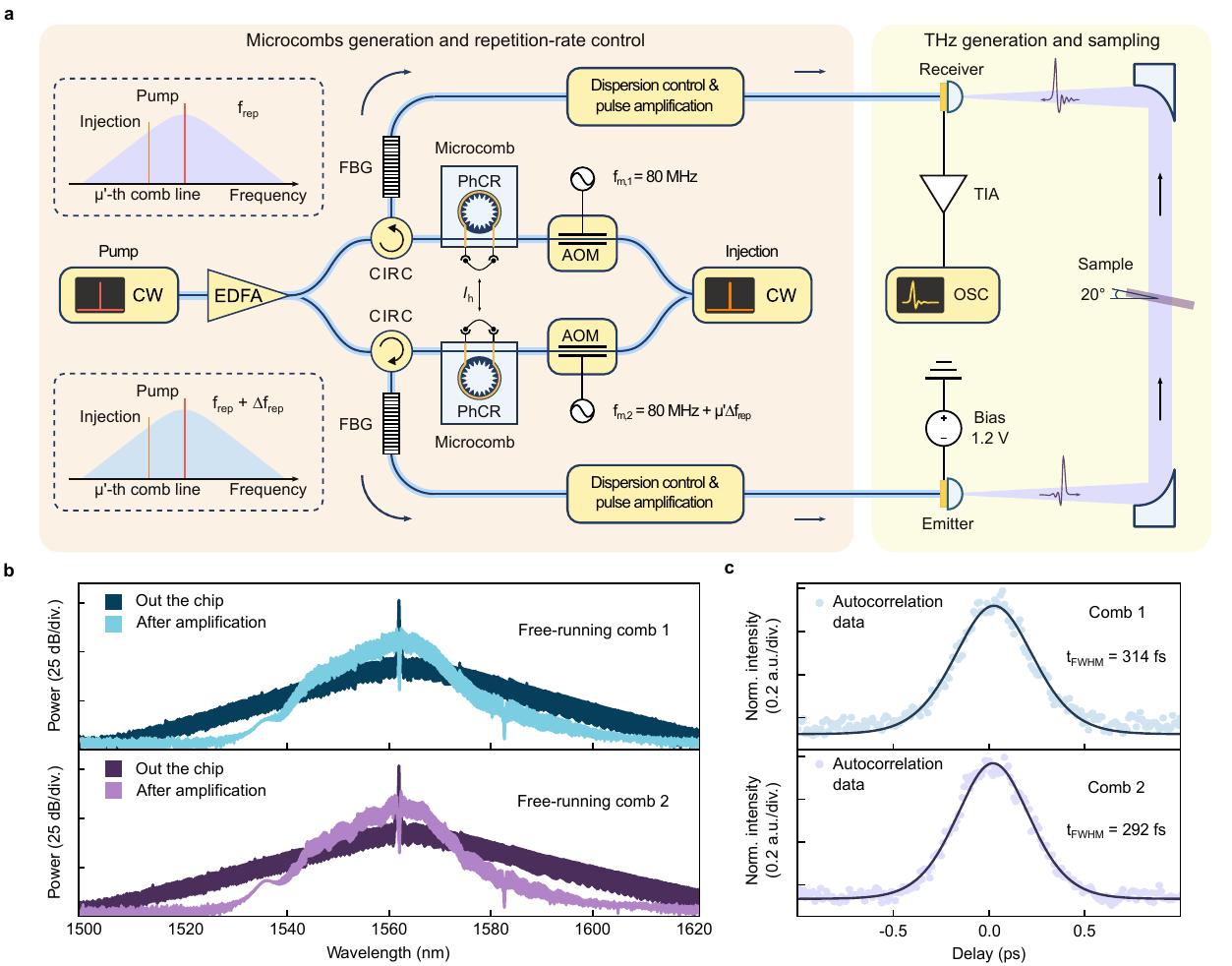}
    \caption{\textbf{Experimental setup and optical pulses characterization.}
      \textbf{a}, Experimental setup in two parts: microcomb generation and repetition rate control, as well as, THz generation and sampling. CW continuous wave lasers for pump and injection; AOM, acousto-optic modulator; EDFA, erbium-doped fibre amplifier; CIRC, circulator; FBG, fibre Bragg grating; TIA, transimpedance amplifier; OSC, oscilloscope. EDFAs and waveshaper are used for pulse amplification and compensation of dispersion. Arrows indicate the direction of light propagation. 
      \textbf{b}, Optical spectra of free-running combs before (dark blue and dark lilac) and after the amplification and dispersion compensation (light blue and light lilac), where we also filter out the pump and injection modes (dips at 1561~nm and 1582~nm respectively). 
      \textbf{c}, Second-harmonic intensity autocorrelation measurements after amplification and dispersion compensation yield pulse durations of $\sim$~300~fs, assuming $\text{sech}^2$-pulses. 
      }
    \label{fig:setup}
\end{figure*}

\subsection*{Results}

\begin{figure*}[!t]
  \centering
  \includegraphics[width=\textwidth]{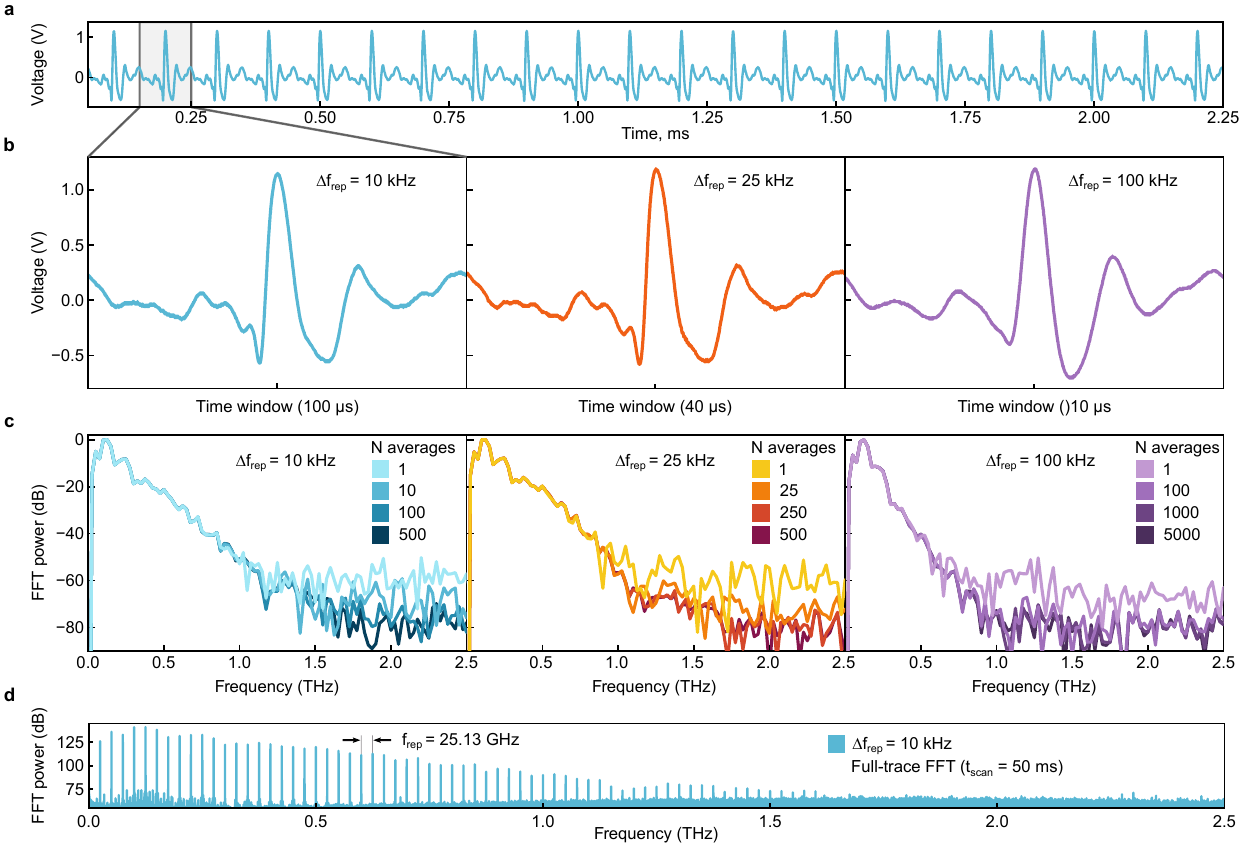}
  \caption{\textbf{THz generation and detection.}
    \textbf{a}, Recorded THz sampling signal with $\Delta f_\mathrm{rep}$ = 10~kHz.
    \textbf{b}, THz transients sampled over the duration of a single delay scan $1/\Delta f_\mathrm{rep}$ for different values of $f_\mathrm{rep}$ = 10, 25, 100~kHz.
    \textbf{c}, Power spectra (normalized to the peak) obtained from a single delay scan and with averaging $N$ delay scans with an effective acquisition time of $N / \Delta f_\text{rep}$ for $f_\mathrm{rep}$ = 10, 25, 100~kHz. 
    \textbf{d}, Power spectrum of the full acquired trace for $\Delta f_\mathrm{rep}$ = 10~kHz for 50~ms averaging. The frequency components' spacing of the \textit{THz comb} correspond to the microcomb repetition-rate $f_\mathrm{rep}$ = 25.13~GHz.
  }
    \label{fig:exp}
\end{figure*}

The setup (\reffig{fig:setup}{a}) consists of two blocks: a dual-microcomb source providing two amplified 25~GHz femtosecond pulse trains with slightly detuned repetition-rates, and the THz-TDS part comprising the THz-generating and sampling PCAs. The microcombs are based on two Si$_3$N$_4$ racetrack photonic crystal microresonators (PhCR) with free spectral ranges close to 25~GHz, both pumped by a common CW laser of frequency $f_\mathrm{p}$ (1560~nm wavelength) delivering an on-chip power of 60~mW in each resonator. A corrugation along part of the resonator waveguide shapes the dispersion such that single dissipative Kerr soliton (DKS) states form deterministically. In this configuration, the DKS pulses form in the blue-detuned regime and counter-propagate with respect to the pump (see Methods)~\cite{ulanov2026brightsolitonshybriddispersionphotonic}; they are out-coupled via circulators. Free-running, the repetition-rates of the two microcombs differ by approximately 300~kHz.

To enable longer acquisition and signal averaging, we stabilize the microcombs' repetition rate via sideband injection locking~\cite{moille2023KerrinducedSynchronizationCavity,wildi2023SidebandInjectionLocking,ulanov2025AttosecondlevelSynchronisationChipintegrated}: a second CW laser of frequency $f_\mathrm{inj}$ (1582~nm wavelength, on-chip power $\sim5$~mW) is injected into each resonator (co-propagating with the solitons), at comb mode $\mu = 100$ relative to the pump. Via the Kerr nonlinearity, the comb line at mode $\mu$ locks to the injected frequency, which fixes the repetition-rate to $f_\mathrm{rep} = |f_\mathrm{inj} - f_\mathrm{p}|/\mu$. Prior to injection, the laser is split into two paths where it is differentially frequency-shifted by acousto-optic modulators (AOMs) driven at $f_{\mathrm{m},i} \approx 80$~MHz (microcomb index $i = 1,2$), so that the two microcombs lock to $f_{\mathrm{rep},i} = |f_\mathrm{inj} + f_{\mathrm{m},i} - f_\mathrm{p}|/\mu$. Their difference, $\Delta f_\mathrm{rep} = |f_{\mathrm{m},2} - f_{\mathrm{m},1}|/\mu$, is therefore set purely electronically via the AOM modulation frequencies. The values of $\Delta f_\mathrm{rep}$ of 10 to 100~kHz used here correspond to AOM frequency difference of 1 to 10~MHz. Both AOM drivers share a common clock. 

At each microcomb output, a fibre Bragg grating removes the residual pump light, an erbium-doped fibre amplifier (EDFA) amplifies the pulses to up to 200~mW of average power (8~pJ pulse energy). A programmable spectral filter (waveshaper) compensates the dispersion accumulated in the fibre path and removes the injection laser line. The optical spectra of the two free-running microcombs and the amplified spectra are shown in \reffig{fig:setup}{b}. Second-harmonic intensity autocorrelation measurements of the amplified pulses indicate near transform limited pulse durations of approximately 300~fs (FWHM, assuming a sech$^2$ pulse shape), corresponding to a peak power of approximately 24~W (\reffig{fig:setup}{c}).

The THz-TDS part of the setup consists of two fibre-coupled, photoconductive modules (see Methods) with hyperhemispherical silicon lenses. One of the two 25~GHz pulse trains (average power 20~mW) is sent to the emitter PIN photodiode, which is biased at $-1.2$~V. At the PIN photodiode each optical pulse generates a THz transient, resulting in a train of THz pulses with 25~GHz repetition-rate. The THz beam is collimated by a parabolic mirror, propagates in air, and is focused by a second parabolic mirror onto the unbiased receiver PCA. There, the second pulse train (average power 200~mW) gates the receiver, such that a photocurrent proportional to the momentary THz field is generated. The current is converted into a voltage by a transimpedance amplifier (TIA) and recorded by an oscilloscope. As the two microcomb repetition-rates differ by $\Delta f_\mathrm{rep}$, the sampling gate slides temporally through the THz transient, sweeping across the full delay window of $1/f_\mathrm{rep} = 40$~ps once every $1/\Delta f_\mathrm{rep}$ (100~$\upmu$s at $\Delta f_\mathrm{rep} = 10$~kHz). The recorded signal corresponds to the THz waveform stretched in time by $f_\mathrm{rep}/\Delta f_\mathrm{rep}$ ($2.5\times10^{6}$ at 10~kHz), and is recorded by an oscilloscope (as shown in \reffig{fig:concept}{a}). A trigger signal at $\Delta f_\mathrm{rep}$, derived from the same clock as the AOM drivers, provides the timing reference for the recorded traces and allows consecutive scans to be aligned and averaged.

An example of the recorded THz signal is shown in \reffig{fig:exp}{a} for $\Delta f_\mathrm{rep} = 10$~kHz; the TIA bandwidth and gain are set to 1.8~MHz and $10^6$~V/A. The sampled THz transient repeats every $1/\Delta f_\mathrm{rep} = 100$~$\upmu$s. Owing to the 25~GHz repetition-rate, the dead-time between consecutive transients is short, in contrast to a typical sub-percent duty cycle of conventional ASOPS systems with $\sim$~100~MHz repetition-rate. \reffig{fig:exp}{b} shows individual delay scans, each of duration $1/\Delta f_\mathrm{rep}$, for $\Delta f_\mathrm{rep}$ = 10, 25 and 100~kHz; the transient is clearly resolved in all cases. Owing to the stabilized repetition rate, consecutive scans can directly be coherently averaged over $N$ scans to improve the signal-to-noise ratio.

\begin{figure*}[!t]
  \centering
  \includegraphics[width=\textwidth]{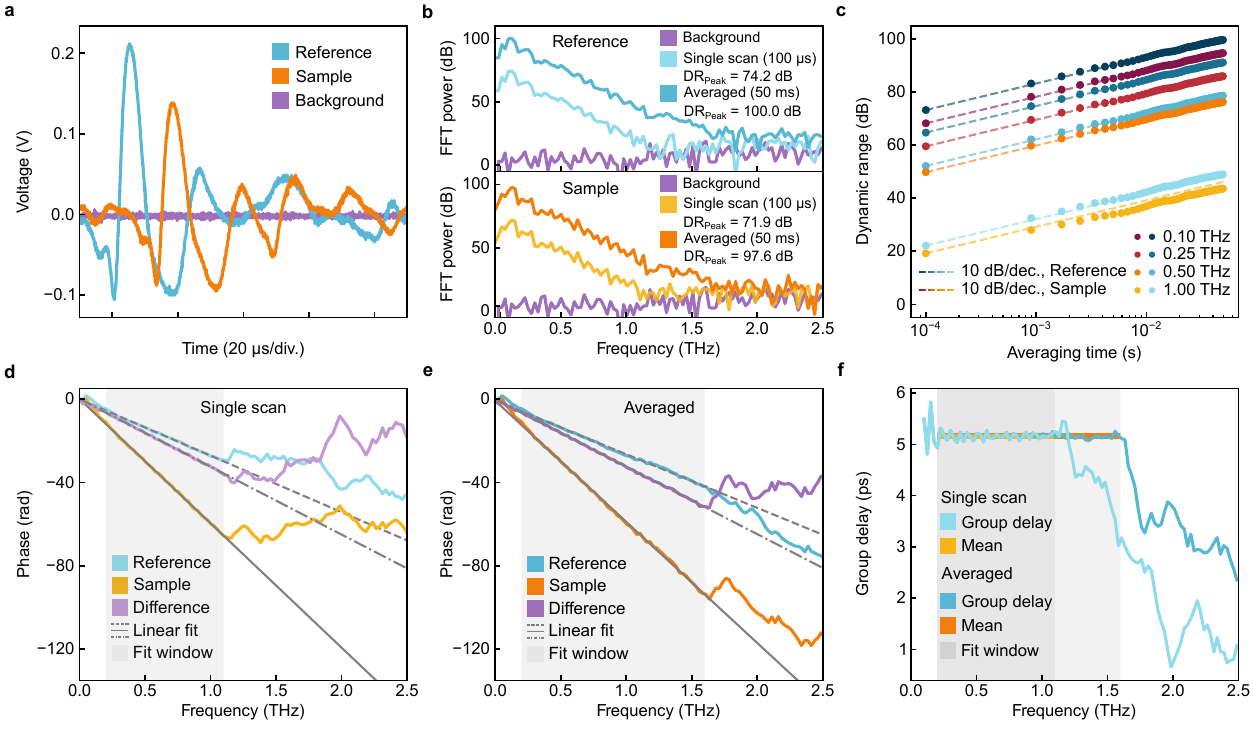}
  \caption{\textbf{THz spectroscopy of a silicon wafer.}
    \textbf{a}, Single delay scan time-domain traces with $\Delta f_\text{rep}$ = 10~kHz for a Si wafer sample (orange), reference (blue, no wafer), and background (lilac, THz beam blocked) signal. 
    \textbf{b}, Power spectra of the time domain signals without the wafer (top panel) and with wafer sample (bottom panel) measurement (normalized to the noise level, extracted as the root-mean-square of the background measurement, lilac). Power spectra are shown for a single delay scans (light blue and yellow) as well for $N=500$ averages over 50~ms of acquisition time (blue and orange).
    \textbf{c}, Dynamic range for 0.1, 0.25, 0.5 and 1~THz frequencies as a function of the averaging time, showing a linear scaling~\cite{Liebermeister2021optoelectronicthztds}.
    \textbf{d}, Unwrapped phase information obtained from a single delay scan for reference (without wafer) and sample (with wafer) signal after background subtraction; we also indicate their linear phase difference encoding the group delay of THz signal by the wafer~\cite{Jepsen2019PhaseRetrieval}. 
    \textbf{e}, Same as panel d, but for the averaged data ($N=500$).
    \textbf{f}, Additional group delay, imposed by the sample as derived from the differential phase slope between reference and sample measurements (see main text).
  }
    \label{fig:sample}
\end{figure*}

To assess the THz bandwidth, we Fast-Fourier transform (FFT) the recorded delay scans, considering both single scan ($N = 1$) and coherent averages over $N$ scans, corresponding to scan times of $N/\Delta f_\mathrm{rep}$ (\reffig{fig:exp}{c}). The spectra are sampled at the 25~GHz line spacing of the microcombs' repetition-rate, which sets the spectral resolution. It is visible in \reffig{fig:exp}{d}, where the FFT is performed over the full scan of 50~ms. For $\Delta f_\mathrm{rep}$ = 10~kHz, the spectrum extends to $\sim1.6$~THz before reaching the noise floor approximately 80~dB below the maximum. At higher scan rates, the bandwidth is reduced: since the THz waveform is stretched in time by $f_\mathrm{rep}/\Delta f_\mathrm{rep}$, the TIA bandwidth $B_\mathrm{TIA} = 1.8$~MHz limits the detectable THz frequencies to $B_\mathrm{TIA} \times f_\mathrm{rep}/\Delta f_\mathrm{rep}$, i.e., 4.5, 1.8 and 0.45~THz for $\Delta f_\mathrm{rep}$ = 10, 25 and 100~kHz, respectively. 
The bandwidth observed at 10~kHz is thus not limited by the TIA but by the antennas and the optical pulse duration. Higher scan rates, up to the MHz range supported by the microcombs, would require a TIA of correspondingly higher bandwidth, at the expense of transimpedance gain. Signal averaging over multiple scans lowers the noise floor, but the improvement saturates for large $N$, as contributions to the background that are synchronous with the THz waveform, such as electronic pickup at harmonics of $\Delta f_\mathrm{rep}$, average coherently and, therefore, are not reduced. Unlike incoherent noise, however, this coherent background can be measured with the THz beam blocked and subtracted, as shown below.

Next, we perform a transmission measurement on a 600~$\upmu$m thick high-resistivity float-zone silicon (HRFZ-Si) wafer, whose refractive index $n = 3.4175$ is essentially constant and whose absorption and group velocity dispersion are negligible within the obtained THZ signal~\cite{Dai2004hrfzref}. The wafer is placed in the collimated THz beam, slightly tilted at an angle of approximately 20$^\circ$ to avoid back-reflections (for simplicity we ignore this tilt in the considerations presented below). The TIA is set to $10^5$~V/A gain and 3.5~MHz bandwidth. At $\Delta f_\mathrm{rep}$ = 10~kHz, we record the signal with the wafer in the beam path (sample), without it (reference), and with the beam blocked (background). Single scan signals are shown in \reffig{fig:sample}{a}; the dominant effect of the wafer is a delay of the transient by approximately 5~ps, corresponding to the additional propagation delay $(n-1)d/c$, together with a reduced amplitude due to Fresnel reflection losses. In addition, a weaker replica of the transient, delayed by the wafer-internal round-trip time $2nd/c \approx 14$~ps, results from the Fabry–P\'erot reflections within the wafer.

We process sample, reference, and background traces as described above, for single delay scans and for coherently averaged scans, and subtract the background signal in the time domain prior to the FFT operation to remove the coherent background. The resulting reference and sample spectra, normalized to the noise-level extracted as the root-mean-square of the background measurement, are shown in \reffig{fig:sample}{b}, along with the background spectrum in the top and bottom panels, respectively. The maximal dynamic range (peak-to-noise-floor ratio) for reference (sample) reaches 74~dB (72~dB) in a single scan of 100~$\upmu$s in duration, and 100~dB (98~dB) after $\sim$~50~ms of coherent averaging of $N=500$ scans. 
To characterize how the dynamic range scales with measurement time, we evaluate it at several THz frequencies (0.1, 0.25, 0.5 and 1~THz) as a function of the averaging time $N/\Delta f_\mathrm{rep}$ (\reffig{fig:sample}{c}). At all frequencies the dynamic range increases by 10~dB per decade of averaging time, i.e., from 60~dB in a single 100~$\upmu$s scan to 87~dB after 50~ms at 0.25~THz, as expected for coherent averaging~\cite{Liebermeister2021optoelectronicthztds}, where the signal power grows as $N^2$ while the noise power grows only as $N$.
Apart from the Fresnel reflection losses and the etalon fringes, the sample and reference spectra show no significant difference in spectral amplitude, consistent with the negligible absorption of HRFZ-Si. However, the spectral phase is modified by the wafer: the difference $\Delta\varphi(\omega)$ between the unwrapped sample and reference phases (\reffig{fig:sample}{d, e}) is linear in (angular) frequency $\omega$, corresponding to an additional spectrally flat group delay $\tau = \Delta\varphi(\omega)/\omega =\mathrm{const}$ imposed by the wafer on the THz wave. 
From the single scan and coherently averaged measurements we derive for each frequency the group delay as shown in \reffig{fig:sample}{f}, obtaining mean (additional) group delays of 5.15~ps and 5.16~ps for the two measurements over the bandwidth from 0.2 to 1~THZ and 1.6~THz, respectively. Assuming  $n = 3.4175$, we estimate the thickness of the sample from the measured additional group delay as $d\approx \tau c/(n-1) = 639.3~\upmu$m and 639.5~$\upmu$m, respectively. The good agreement between both values demonstrates that a single delay scan suffices for rapid and precise thickness measurements.

\subsection*{Discussion}
In conclusion, we have demonstrated asynchronous THz-field sampling with two chip-integrated 25~GHz soliton microcombs. The setup does neither require mechanically moving parts, nor active feedback loops. In contrast to conventional ASOPS, the high repetition rate reduces the dead time when measuring fast condensed-matter THz transients, and enables rapid acquisition of THz waveforms and spectra. The acquisition rates of up to 100~kHz demonstrated here are limited by the transimpedance amplifier and not by the microcomb source, which would support MHz rates. Kerr-nonlinear synchronization of the two microcombs via sideband injection locking provides the timing stability required for coherent averaging. The trade-off of THz-TDS with high 25 GHz repetition-rate is the reduced spectral resolution given by the repetition-rate, which suits solids, thin films, and liquids with broad spectral features but not gas-phase spectroscopy of narrow rotational lines. 

While pulse amplification and dispersion compensation remain off-chip in the present demonstration, recent progress in integrated pulse amplifiers~\cite{gaafar2024FemtosecondPulseAmplification, zeng2026IntegratedYtterbiumGain}, on-chip dispersion compensation~\cite{tan2010MonolithicNonlinearPulse,sinobad2025DispersionCompensatingSilicon, sinobad2025DispersionCompensatingSilicon}, waveguide integrated photoconductive elements\cite{WaveguideIntegratedPIN}, and electro-optic generation and detection in thin-film lithium niobate~\cite{gaier2025wirelessmillimeterwaveelectroopticsfilm} point towards fully integrated systems for rapid THz-field sampling~\cite{Rajabali2023presentfutureTHZ}.

\subsection*{Methods}
\small

\paragraph{Microresonators characterization.} 
Two photonic crystal microresonators (PhCR) are commercially manufactured on a Si$_3$N$_4$ photonic chip platform by ultraviolet stepper optical lithography by LIGENTEC SA. The microresonators share an identical racetrack geometry, yielding a free spectral range (FSR) of 25.132 GHz. Their waveguide cross-section of $2\times0.8~\mu$m on an 800-nm-thick platform provides anomalous group-velocity dispersion ($D_2/2\pi = 59$ kHz) with a local region of $D_2'/2\pi = 20$ MHz near the pump due to engineered mode hybridization~\cite{ulanov2026brightsolitonshybriddispersionphotonic}. Each microresonator is critically coupled with loaded quality factor of $Q \approx 5\times10^6$ corresponding to a total resonance linewidth of $\kappa/2\pi \approx 40$ MHz.

\paragraph{THz antennas.}
Two fiber-coupled optoelectronic THz antennas are used, one as an emitter and one as a receiver. The emitter is a waveguide-integrated PIN photodiode with an attached planar antenna based on an indium phosphide platform \cite{deumer2024ultra}. It is reversibly biased with $-1.2$~V. 
The receiver is a rhodium-doped InGaAs photoconductive antenna (PCA)~\cite{kohlhaas2019photoconductive}. 
Both emitter and receiver are mounted on hyper-hemispherical silicon lenses and integrated into fiber-pigtailed modules. 


\small
\paragraph{Funding.}
\small
This project has received funding from the European Research Council (ERC grant agreement No 853564), the European Innovation Council (EIC grant agreement No 101137000), and the Helmholtz Young Investigators Group VH-NG-1404; the work was supported through the Maxwell computational resources operated at DESY.


\printbibliography

\end{refsection}
\end{document}